\documentclass[prl,twocolumn,tightenlines,nofootinbib]{revtex4}
\usepackage{graphicx}
\usepackage{multirow}
\usepackage{amsmath,amsfonts,amssymb}
\usepackage{bm}
\usepackage{dcolumn}
\usepackage{graphicx}
\usepackage{epsfig}
\usepackage{color}

\begin{document}

\title{Two clock transitions in neutral Yb for the highest sensitivity to variations of the fine-structure constant}

\author{M.~S.~Safronova$^{1,2}$}
\author{S.~G.~Porsev$^{1,3}$}
\author{Christian Sanner$^{4}$}
\author{Jun Ye$^{4}$}

\affiliation{
$^1$Department of Physics and Astronomy, University of Delaware, Newark, Delaware 19716, USA\\
$^2$Joint Quantum Institute, National Institute of Standards and Technology and the University of Maryland,
Gaithersburg, Maryland 20742, USA \\
$^3$Petersburg Nuclear Physics Institute of NRC ``Kurchatov Institute'', Gatchina 188300, Russia\\
$^4$JILA, NIST and University of Colorado, Boulder, Colorado 80309, USA\\
and Department of Physics, University of Colorado, Boulder, Colorado 80309, USA}
\date{\today}

\begin{abstract}
We propose a new frequency standard  based on a $4f^{14} 6s6p~ ^3\!P_0 - 4f^{13} 6s^2 5d ~(J=2)$ transition in neutral Yb. This transition has a potential for high stability and accuracy and the advantage of the highest sensitivity among atomic clocks to variation of the fine-structure constant $\alpha$. We find its dimensionless $\alpha$-variation enhancement factor to be $K=-15$, in comparison to the most sensitive current clock (Yb$^+$ E3, $K=-6$), and it is 18 times larger than in any neutral-atomic clocks (Hg, $K=0.8$). Combined with the unprecedented stability of an optical lattice clock for neutral atoms, this high sensitivity opens new perspectives for searches for ultralight dark matter and for tests of theories beyond the standard model of elementary particles. Moreover, together with the well-established $^1\!S_0 -\, ^3\!P_0$ transition one will have two clock transitions operating in neutral Yb, whose interleaved interrogations may further reduce systematic uncertainties of such clock-comparison experiments.
\end{abstract}

\maketitle

The development of optical atomic clocks has made
over a $\mathrm{factor} \: \mathrm{of} \: 1000$ improvement of precision in less than $15 \: \mathrm{years}$ \cite{LudBoyYe15}.
Many clock applications are enabled by the improved precision and high stabilities: study of many-body physics and quantum simulations \cite{ZhaBisBro14,KolBroBot17}, relativistic geodesy \cite{Flu16}, very long baseline interferometry \cite{NorCle11},
searches for the variation of the fundamental constants \cite{SafBudDem17} and dark matter \cite{DerPos14,Arv15,TilLeeBou15,StaFla15,HeeGueAbg16,WsiMorBob16,RobBleDai17,KalYu17}, tests of the Lorentz invariance \cite{PihGueBai17}, redefinition of the second \cite{BreMilPiz17}, and others. These applications and new ideas, for example the use of atomic clocks for gravitational wave detection \cite{KolPikLan16}, need even more precise clocks.

In this work, we propose a new atomic clock with two different clock transitions in a single atom and the highest sensitivity to the variation of the fundamental fine-structure constant $\alpha$ among all currently operating
optical atomic clocks. In particular, we show that the proposed transition offers highly promising accuracy and stability perspectives and is accessible using well-developed technologies with neutral atoms in optical lattices.

The dual clock operation will profit from common mode suppression of many systematic effects.
One-year spaced measurements of the ratio of the two transition frequencies at the $10^{-18}$ level will lead to uncertainties for $\dot{\alpha} / \alpha$ of $\approx 9 \times 10^{-20} \: \mathrm{per} \: \mathrm{year}$, corresponding to a hundredfold improvement over current limits \cite{Godunetal2014, Huntemannetal2014}.
The full potential of the new transition can be exploited in the context of searches for ultralight scalar dark matter~\cite{Arv15} where one tries to detect $\alpha$-oscillations on all accessible time scales.

In the standard model (SM), all fundamental constants are invariable, but
the dimensionless constants become dynamical in a number of theories beyond the SM and general relativity (GR)~\cite{Uza11}. For example, string theories predict the existence of a scalar field, the dilaton, that couples directly to matter~\cite{TayVen88}. Other theories beyond the SM and GR have been proposed in which fundamental constants become dynamic fields, including discrete quantum gravity~\cite{GamPul03}; loop quantum gravity~\cite{TavYun08}; chameleon models~\cite{KhoWel04}; dark energy models with a non-minimal coupling of a quintessence field~\cite{AveMarNun06}.
Searching for variation
of fundamental constants is also a test of the local position
invariance hypothesis and thus of the equivalence principle \cite{Uza11,Uza15}.

The dependence of atomic and molecular spectra on fundamental constants is used to probe their variations from a distant past. Studies of quasar absorption spectra \cite{WebFlaChu99,WebKinMur11,WhiMur15} indicate that the fine-structure constant may vary on a cosmological space-time scale.

The search for the variation of fundamental constants directly relates to the major unexplained phenomena of our Universe: What is the nature of the dark matter?
Scalar bosonic dark matter (DM) in our Galaxy with mass $m_\textrm{DM} <  1~$eV exhibits coherence and behaves like a wave with an amplitude $\sim \sqrt{\rho_{\textrm{DM}}}/m_{\textrm{DM}}$, where $\rho_{\rm{DM}}=0.3 \: \mathrm{GeV/cm}^3$ is the DM density \cite{Arv15}. The coupling of such DM to the standard model
leads to oscillations of fundamental constants and, therefore, to the oscillation of atomic frequencies detectable with atomic clocks ~\cite{Arv15,TilLeeBou15,HeeGueAbg16}.
Dark matter
objects with large spatial extent, such as stable
topological defects built from light non-SM fields, induce transient changes in fundamental constants that may be detectable with networks of clocks \cite{DerPos14,WsiMorBob16,RobBleDai17,KalYu17}.

\begin{figure}[t]
\includegraphics[scale=0.35]{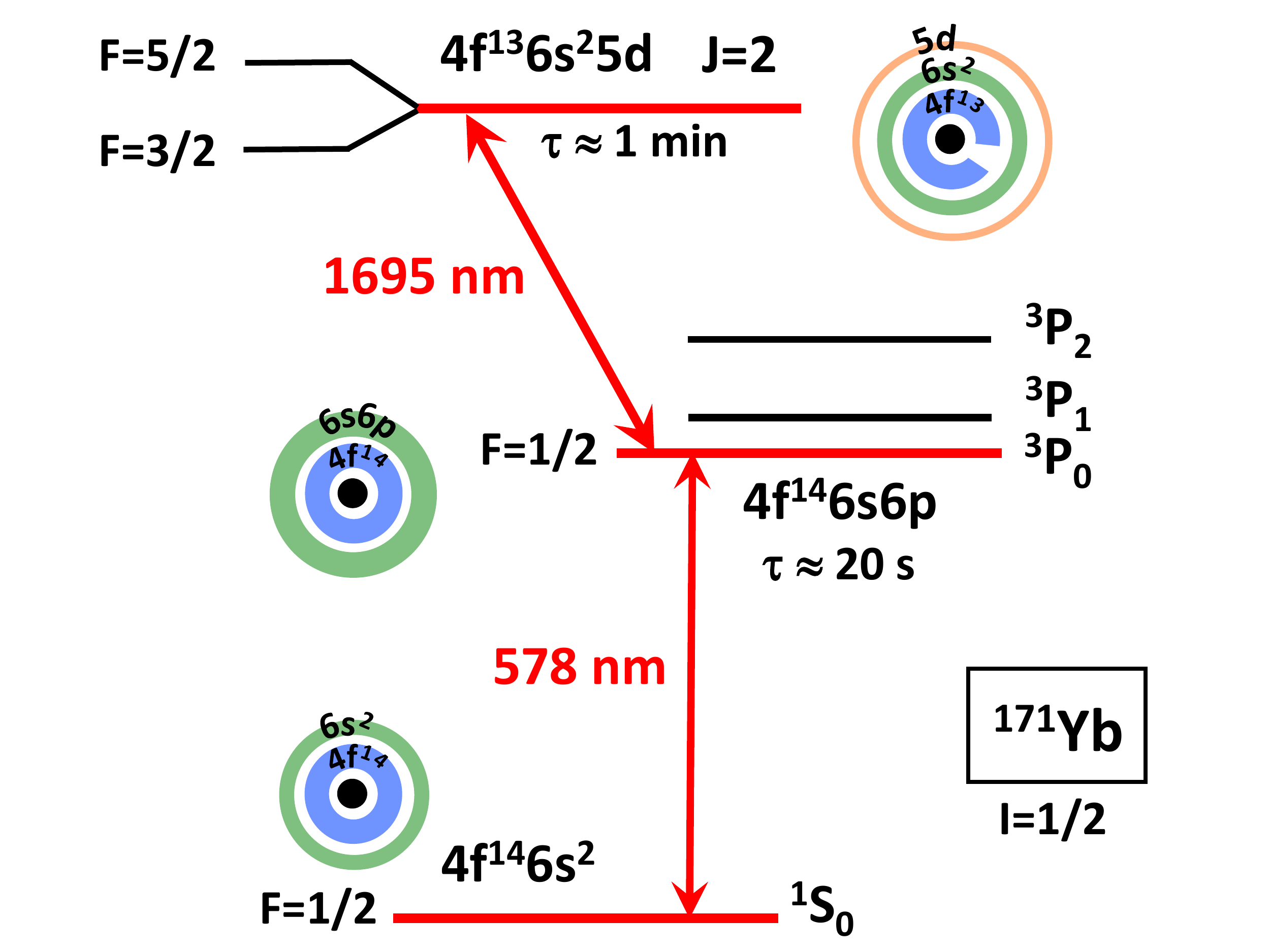}
\caption{Yb energy level scheme illustrating the $4f^{14} 6s^2\, ^1\!S_0$ -- $4f^{14} 6s6p\, ^3\!P_0$ and
$4f^{14} 6s6p\, ^3\!P_0$--$4f^{13} 6s^2 5d\,(J=2)$ clock transitions. The open $4f$ shell of the $J=2$ state leads to the particularly high $\alpha$-sensitivity. Energies are not to scale.}
\label{fig1}
\end{figure}
Clock transition energies $\Delta E$ depend on $\alpha$ if the involved atomic states lead to a nonzero differential sensitivity parameter $\Delta q$ \cite{DzuFlaWeb99,DzuFlaWeb99a} so that
\begin{equation}
\Delta E(\alpha) = \Delta E_0 + \Delta q \left[\left(\frac{\alpha}{\alpha_0}\right)^2-1\right].
\label{eq:wx}
\end{equation}
Here, $\alpha_0$ is the current value of $\alpha$ \cite{CODATA2014},
and $\Delta E_0$ is the transition energy corresponding to $\alpha_0$.
Accordingly, the atomic clock will map small fractional $\alpha \; \mathrm{deviations}$ of any cause or type (temporal, spatial, slow drift, oscillatory, gravity-potential dependent, transient or other) to fractional frequency deviations
\begin{equation}
\frac{\Delta E - \Delta E_{0}}{\Delta E_0} = K \frac{\alpha - \alpha_0}{\alpha_0}
\end{equation}
via the dimensionless enhancement factor $K = 2 \Delta q / \Delta E$. Experimentally, one can detect the variation of $\alpha$ by monitoring the ratio of two clock frequencies with different values of $K$. The specific measurement protocol depends on the type of the $\alpha$-variation, but using clocks with the best stability, total systematic uncertainty, and the highest possible
values of $\Delta K=K_1-K_2$ for clocks 1 and 2 has the highest discovery potential.

There are two types of optical atomic clocks at the present time, based on neutral atoms in optical lattices and based on a single trapped ion. Similar uncertainties have been reached for both kinds,
$2.1\times 10^{-18}$ for a Sr neutral atom clock \cite{NicCamHut15} and $3.2\times 10^{-18}$ for a Yb$^+$ trapped ion clock \cite{HunSanLip16} operating on the electric octupole (E3) transition. A large number (a few 1000s) of simultaneously interrogated atoms leads to much better stability of the neutral atom clocks in comparison to the single ion clock. A record frequency precision  of $2.5\times 10^{-19}$ at 6 hours  averaging time has been just demonstrated with the Sr optical lattice clock at JILA~\cite{MarHutGob17}. The number of atoms in a lattice clock may be significantly increased in a three-dimensional clock~\cite{CamHutMar17}. However, Sr, Yb and Hg lattice clocks have $K$=0.06, 0.37, and 0.8, respectively~\cite{DzuFla09}. Among all currently operating clocks, the $\mathrm{Yb}^+ \; \mathrm{E3}$ transition $4f^{14} 6s\,\, ^2\!S_{1/2} - 4f^{13} 6s^2$~$^2\!F_{7/2}$ has the highest enhancement factor $K=-6$ \cite{DzuFla09}. Comparing it to the E2 transition $4f^{14} 6s\,\, ^2\!S_{1/2} - 4f^{14} 5d$~$^2\!D_{3/2}$ in the same ion yields a clock system with $\Delta K = -7$ \cite{Godunetal2014}. On the other hand, the corresponding $\mathrm{Yb}^+ \; \mathrm{E3}$ clock stability at $6 \; \mathrm{hours}$ is $3.4\times 10^{-17}$~\cite{HunSanLip16}, and two orders of magnitude improvement would be difficult and require realization of the clock with ion chains \cite{KelKalBur17}.

\begin{table} [t]
\caption{Experimental energies~\cite{RalKraRea11} and $\alpha$-variation sensitivity
coefficients $\Delta q$ (in $\mathrm{cm}^{-1}$) for low-lying states of Yb. All values are counted from the ground state, except for the last row, where the energy is given with respect to the metastable $^3\!P_0$ state. $K$ is the dimensionless $\alpha$-variation enhancement factor, $K=2\Delta q/\Delta E$.}
\begin{ruledtabular}
\begin{tabular}{lcccc}
\multicolumn{1}{c}{Level} &
\multicolumn{1}{c}{Term} &
\multicolumn{1}{c}{$\Delta q$}&
\multicolumn{1}{c}{$\Delta E$}&
\multicolumn{1}{c}{$K$}\\
\hline \\[-0.8pc]
$4f^{14} 6s^2$   &$^1\!S_0$&      0 &   0   &     \\[0.3pc]
$4f^{14} 6s6p$   &$^3\!P_0$&  3185 & 17288 & 0.37\\
                 &$^3\!P_1$&  3992 & 17992 &     \\
                 &$^3\!P_2$&  5818 & 19710 &     \\[0.3pc]
$4f^{14} 6s5d$   &$^3\!D_1$&  7878  & 24489 &     \\[0.5pc]
$4f^{13} 6s^25d$ & $J=2$   & -40345 & 23189 &     \\
                 & $J=5$   & -40978 & 25860 &     \\
                 & $J=6$   & -39528 & 27314 &     \\
                 & $J=3$   & -40981 & 27445 &     \\[0.3pc]
$4f^{13} 6s^25d$ & $J=2$   &        &       &     \\[-0.9pc]
                 &         & -43530\footnotemark & 5901$^{a}$  & -15$^{a}$\\
\end{tabular}
\end{ruledtabular}
\footnotetext{Relative to the $4f^{14}6s6p~^3\!P_0$ state.}.
\label{tab1}
\end{table}

\paragraph{Yb two-clock proposal.} The $4f^{14} 6s^2\,\,^1\!S_0 - 4f^{14} 6s6p\,\, ^3\!P_0$ transition in neutral Yb already serves as a basis for a highly accurate frequency standard \cite{SchBroMcG17,BroPhiBel17}. We find that neutral Yb,
being a f-block element,
has another transition,
$4f^{14} 6s6p~ ^3\!P_0 - 4f^{13} 6s^2 5d ~(J=2)$
at an easily accessible wavelength of 1695 nm, that is
suitable for the development of another frequency standard in this atom. The excited state electronic configuration with its open $f$-shell
and the single 5$d$ electron resembles those encountered in the aforementioned $\mathrm{Yb}^+ \; \mathrm{E2}$ and E3 transitions, and
therefore, combining the effects of a pronounced relativistic energy shift \cite{DzubaFlambaumPRA2008} and a smaller transition energy,
we find $K=-15$, a $\mathrm{factor} \: \mathrm{of} \: 18$ higher than for any other lattice clock.
We propose interleaving interrogation of the two Yb clock transitions to reduce the
systematic uncertainties of such clock-comparison experiments. The Yb energy level scheme illustrating both clock transitions is shown in Fig.~\ref{fig1}.
We use the fermionic $^{171}$Yb isotope with  $I=1/2$ for illustration, but the $^{173}$Yb isotope with $I=5/2$ may be used as well.

Since the two clock transitions are expected to have different magic wavelengths (see below), we propose a sequential operation of the two clock transitions, i.e. making measurements with the ``traditional'' $^1\!S_0-\,^3\!P_0$ clock at one cycle and then switching to the second clock transition at the next cycle. All the measurements will be performed at the same spatial location under common electromagnetic and gravitational fields. For the $^3\!P_0 \rightarrow J=2$ clock transition, we start with atoms prepared in the $^1S_0$ ground state in the traditional Yb magic wavelength $\lambda_{\textrm{magic}}=795.36$~nm \cite{LemLudBar09} lattice and use a $\pi$-pulse to drive the population to the $^3\!P_0$ state. As a next step, we adiabatically switch between the two magic wavelength lattices by gradually turning down the intensity of the original lattice while turning up the intensity of the second lattice. Direct
optical pumping assisted preparation \cite{LeTargat2006}
of the $^3\!P_0$ atoms
in the second magic wavelength lattice is also possible. When driving the $^3\!P_0 \rightarrow J=2$ clock transition, the population of the $^3\!P_0$ state is monitored  by using a short-duration $\pi$-pulse to transfer it down to $^1S_0$ and collect laser fluorescence. Below, we discuss relevant properties of the $J=2$ clock state and systematic uncertainties specific to operating the lattice clock with
the $J\neq 0$ level.

\paragraph{Sensitivity to $\alpha$-variation.}
Table~\ref{tab1} gives the experimental energies of the low-lying Yb levels and the calculated $\Delta q$ coefficients, counted from the ground state. The calculations are carried out using the configuration interaction (CI) method treating Yb as a system with 16 valence electrons.
Computational details are described in~\cite{suppl}.
The change $\delta K$ between the different CI sets for the proposed clock was less than 0.1\%.
To estimate the uncertainty of the $q$ coefficients, we also carried out much simpler Dirac-Hartree-Fock calculation with a single non-relativistic configuration, and find only 5\% change in the value of $\Delta q$ for the $4f^{14} 6s6p~ ^3\!P_0 - 4f^{13} 6s^2 5d ~(J=2)$ transition despite a drastic difference in the transition energy ($\sim 10000 \; \mathrm{cm}^{-1}$), confirming that the values of $q$ depend weakly on the treatment of the electronic correlations.

\paragraph{Upper clock state lifetime and decay channels.}
The decay of $4f^{13} 6s^2 5d ~(J=2)$ to the ground state can occur via the magnetic quadrupole or electric octupole transition and is negligibly weak. Therefore, the $J=2$ level decays via the magnetic-dipole ($M1$) and electric-quadrupole ($E2$) transitions to the odd $4f^{14} 6s6p ~^3\!P_J$ levels. The main decay channel of the $4f^{13} 6s^2 5d ~(J=2)$ state is the $M1$
$(J=2) -\, ^3\!P_1$ transition \cite{Dzuba}, which is confirmed by our CI calculations.
 The calculation of the $M1$ and $E2$ transition amplitudes is complicated by the mixing of the $J=2$ and
$4f^{14} 6s6p ~^3\!P_2$ states.
The \textit{ab initio} CI calculation of the matrix elements does not correctly reproduce this mixing. A model computation~\cite{suppl} aimed at the proper description of the level mixing yields a lower lifetime bound
of $\sim 1$ min. The branching ratio to the $^3\!P_0$ level is  $\sim 3-5\%$ and the clock transition can
be driven with a direct laser excitation from the metastable $^3\!P_0$ level.
The radiative decays from the $J=2$ to the $^3\!P_1$ and $^3\!P_2$ levels are irrelevant at normal time scales for operating clock cycles. In fact, the clock transition linewidth is limited by the lifetime of $^3\!P_0$, just as in the traditional $^1\!S_0 - ^3\!P_0$ transition.

\paragraph{Zeeman shifts.}
The $J=2$ state is more sensitive to magnetic field fluctuations than the $^1\!S_0 - ^3\!P_0$ transition.
This increased sensitivity and the vector and tensor light shifts described below are the two major experimental challenges when trying to exploit the full potential of the new clock transition. Therefore, we discuss in detail various technical requirements and possible experimental strategies.
Regarding the $B$ field sensitivity, we propose to drive two $\pi$-transitions from $^3\!P_0$, $m_F =  \pm 1/2$ states to $J=2$, $F=3/2$, $m_F = \pm 1/2$ states, respectively. The sum of these two transition frequencies is field insensitive to first order. The difference, on the other hand, will provide a measurement of the $B$ field and its potential fluctuation.
Particularly,
we can first use the traditional
$^1\!S_0 -\, ^3\!P_0$ clock transition to null out the residual field to the level of $1 \: \mathrm{mGauss}$. After applying a bias $B$ field of for example $10 \: \mathrm{mGauss}$, we can use the difference of the two $^3\!P_0 \rightarrow J=2$ $\pi$ transitions to enable a magnetic field servo during the clock operation~\cite{BloomNature14}. Having a well-defined quantization $B$ axis will be very important for the precise control of the lattice vector and tensor AC Stark shift.

The frequency separation of the two $\pi$ transitions will also allow us to determine the accurate value of the bias $B$ field for the evaluation of the second order Zeeman effect.
The frequency shift by a magnetic field is proportional to the electronic magnetic moment for the $J=2$ state instead of the nuclear magnetic moment for the $^3\!P_0$ clock state. Therefore, one can use a much smaller $B$ field (than that used in a conventional lattice clock) to bias the two $\pi$ transitions apart for the clock operation.
This also implies that second order Zeeman shifts are kept at correspondingly small values.
If residual magnetic field noise limits the attainable coherence time on the $^3\!P_0 \rightarrow J=2$ transition one can devise a
synchronous
version of the above field noise cancelation scheme by driving the respective transitions of opposite sensitivity simultaneously and performing differential population measurements.

\paragraph{Optical lattice Stark shifts: magic wavelengths.}
In the neutral atom optical clock, the atoms are trapped in an optical lattice operating at the magic wavelength
at which the ac Stark shift of a clock transition  is minimized to a high level of precision \cite{KatIdoKuw99,YeVerKim99}.
In the main approximation, the ac Stark shift is determined by the frequency-dependent electric dipole polarizabilities of the clock states \cite{MitSafCla10}. Therefore, the magic wavelengths can be determined by finding the frequencies
where the ac polarizabilities of the two clocks states are the same.
The total polarizability of a state $|JM\rangle$ is given by
$$
\alpha=\alpha_0+\alpha_2 \frac{3M^2-J(J+1)}{J(2J-1)},
$$
where $J$ is the total angular momentum and $M$ is the corresponding magnetic quantum number.
The scalar polarizability, dominated by the contribution of the valence electrons, may be expressed as the
sum over intermediate $k$ states allowed by the electric-dipole selection rules~\cite{MitSafCla10}
\begin{equation}
\alpha_{0}(\omega)=\frac{2}{3(2J+1)}\sum_k\frac{{\left\langle J_k\left\|D\right\|J\right\rangle}^2(E_k-E_J)}
{(E_k-E_J)^2-\omega^2},
\label{eq-1}
\end{equation}
where the frequency $\omega$ is assumed to be at least several linewidths off
resonance with the corresponding transitions and ${\left\langle J_k\left\|D\right\|J\right\rangle}$ are the
reduced electric-dipole matrix elements. Linear polarization
is assumed. The expression for the tensor polarizability has a similar structure.

The scalar polarizability of the $^3\!P_0$ state can be calculated with a few percent accuracy using the CI+all-order
method \cite{SafKozJoh09} as described in \cite{SafPorCla12}.  There is no tensor contribution to the $^3\!P_0$ polarizability for $^{171}$Yb with $I=1/2$.
The $J=2$ state cannot be treated with the CI+all-order method due to the presence of the $4f$  hole in the electronic
configuration, but the resonant structure of Eq.~({\ref{eq-1}}) allows to estimate the behavior of the $4f^{13} 6s^2 5d\, (J=2)$ polarizability to predict the presence of the magic wavelengths with the $^3\!P_0$ state.
We expect that the upper clock state $4f^{13} 6s^2 5d ~(J=2)$ has sufficiently strong E1 transitions to the $4f^{13} 6s^2 6p ~(J=2,3)$ states since they involve direct one-electron $5d_{3/2}-6p$ transitions. The resonant wavelengths for a number of such transitions are listed in Table~\ref{tab3}, together with the $E1$ transitions for the $^3\!P_0$ state in this wavelength region.
\begin{table} [t]
\caption{\label{tab3} Resonant wavelengths $\lambda$ (in nm) corresponding to the $E1$ transitions contributing to the polarizabilities of the $4f^{13} 6s^2 5d ~(J=2)$ and $^3\!P_0$ clock states. }
\begin{ruledtabular}
\begin{tabular}{cc}
\multicolumn{1}{c}{Transition} &
\multicolumn{1}{c}{$\lambda$}\\
\hline \\[-0.8pc]
$4f^{13}6s^25d~ (J=2) - 4f^{13}6s^26p_{1/2} ~(J=3)$&  	1127  \\
$4f^{13}6s^25d~ (J=2) - 4f^{13}6s^26p_{3/2} ~(J=2)$&  	 833   \\
$4f^{13}6s^25d~ (J=2) - 4f^{13}6s^26p_{3/2} ~(J=3)$&  	 792   \\[0.5pc]
$4f^{14}6s6p~ ^3\!P_0 - 4f^{14} 5d6s~^3\!D_1$      &    1389 \\
$4f^{14}6s6p~ ^3\!P_0 - 4f^{14} 6s7s~^3\!S_1$      &     649  \\
\end{tabular}
\end{ruledtabular}
\end{table}

The $^3\!P_0$ polarizability does not have a resonance between 700~nm and 1200~nm, while the $J=2$ clock state has 3 resonances
leading to several polarizability crossings between the two states. For example, there will be a magic wavelength between the 792~nm and 833~nm, where the $J=2$ polarizability curve has to cross the $^3\!P_0$ polarizability, which is slowly varying (from 149 a.u. to 113 a.u.) for this entire interval.
It would be particularly interesting to experimentally locate the magic wavelength below 792~nm to ascertain how close it is to the  $^1\!S_0 -\, ^3\!P_0$ clock magic wavelength of 759.36~nm~\cite{LemLudBar09}. The other $E1$ transitions from the even state  contributing to the $J=2$ polarizabilities potentially lead to more magic wavelengths. The final choice for the magic wavelength for the new clock transition will partly depend on the rate of
coherence-limiting off-resonant single-photon scattering
and partly on the structure of vector and tensor Stark shifts.

\paragraph{Vector and tensor light shift.}

Non-zero electronic angular momentum of the $J=2$ state gives rise to much larger vector and tensor shifts in comparison with the $J=0$ clock states. Therefore, one must ensure that
the lattice light has purely linear polarization,
e.g., by using high quality polarizers inside the vacuum chamber,
and that it is exactly aligned with the quantization axis set by the $B$ field. Otherwise, even if we precisely stabilize the overall intensity of the lattice light, the tensor shift may drift if the lattice polarization wanders.
Furthermore, the clock light should also have purely linear polarization. The vector light shift can be canceled by averaging the two $\pi$ transitions for $m_F = \pm 1/2$.  In terms of tensor shift, $^{171}$Yb has a nuclear spin of 1/2, leading to no tensor shift for $^3\!P_0$, but for $J=2, \: F=3/2, \: m_F = \pm1/2$ we will have a large tensor shift. Hence polarization control is extra important. Alternatively, $^{173}$Yb (with a nuclear spin of 5/2) allows for the $J=2, \: F=1/2$ state with no tensor light shift but leads to $F=5/2$ for $^3\!P_0$, which possesses a small but finite tensor shift. Both isotopes should be considered for further evaluation of the schemes to minimize tensor light shifts.

\paragraph{Other applications of the $J=2$ state.}
Generally, it is advantageous to have in the same atom access to two clock transitions with different sensitivities to various external fields. In addition to precise differential shift measurements and the creation of synthetic clock frequencies \cite{Yudinetal2011} it will be possible to coherently drive the two transitions simultaneously.

The $4f^{13} 6s^2 5d~ (J=2)$ level is not only sensitive to changes of $\alpha$ but also suitable for testing local Lorentz invariance (LLI) \cite{ShaOzeSaf18}. It is possible to use the $J=2$ state to set limits on the Standard Model Extension (SME) parameters quantifying the LLI violation in the electron-photon sector. The LLI test does not require actual clock operation, but rather a monitoring of the Zeeman splitting for the states in the $J=2$ manifold.

A scheme to probe new light force-carriers, with spin-independent
couplings to the electron and the neutron, using  precision isotope shift spectroscopy
was proposed in \cite{BerBudDel17,FruFucPer17}. The method requires to measure two transition frequencies for different electronic states for four isotopes. The bounds on new physics are extracted from limits on the linearity of King plots with minimal theory. Two transitions proposed here are particulary well suited for such a test and provide the only known case of
two such different metastable transitions in a neutral atom clock system. However, the scheme requires bosonic Yb isotopes which have no hypefine mixing to make the $^1\!S_0 -\, ^3\!P_0$
transition weakly allowed. One alternative is to mix the $^3\!P_J$ levels by applying a magnetic field
\cite{LudBoyYe15}
but we expect that such a large field has to be turned off for excitation to the $J=2$ clock state due to the large Zeeman shifts of the $J=2$ level.
Turning strong magnetic fields on and off while keeping them stable may be technically challenging and further investigation is needed to evaluate the measurement accuracy that may be reached for bosonic isotopes.

In summary, we proposed a new clock transition in the Yb atom with the highest sensitivity to the variation of the fine-structure constant among the optical atomic clocks, and, therefore, ultralight dark matter searches. We described a suitable two-clock interrogation scheme and discussed systematic uncertainties associated with the use of the $J\neq0$ level in a neutral atom lattice clock. The proposed scheme may also be used for tests of Lorentz violation and to probe new light force carriers.

\begin{acknowledgments}
We thank Vladimir Dzuba for bringing out attention to the M1 decay channels.
This research was performed in part under the sponsorship of the Office of Naval Research, award number N00014-17-1-2252. We also acknowledge support from NIST and NSF. C.~S. thanks the Humboldt Foundation for support. S.~P. acknowledges support from Russian Foundation for Basic Research under Grant No. 17-02-00216. M.~S. and S.~P. are grateful to JILA for hospitality.\\
\end{acknowledgments}


\onecolumngrid
\begin{center}
\Large{\textbf{Supplemental Material}}\\
\end{center}
\subsection{The calculation of the $q$ coefficients}
To calculate the $q$ coefficients, we have repeated all configuration interaction (CI)
 calculations with the CODATA value of $\alpha$ and modified values $\alpha_{-}=\alpha\sqrt{1-\lambda} $  and   $\alpha_{+}=\alpha\sqrt{1+\lambda} $   with $\lambda=1/8$) to  calculate $dE/d\alpha$. Such calculations were repeated several times under different approximations, using deferent basis sets, and increased sizes of the configuration space to ensure stability of the results.

\subsection{The calculation of the transition rates and branching ratios}
The $J=2$ level decays via the magnetic-dipole ($M1$) and electric-quadrupole ($E2$) transitions to the odd $4f^{14} 6s6p ~^3\!P_J$ levels.
The calculation of the  transition amplitudes requires special care. The main configurations of the $4f^{13} 6s^2 5d ~(J=2)$ and
$4f^{14} 6s6p ~\,^3\!P_1$ states differ by two electrons.
The $M1$ transition between these states occurs only due to admixtures of other configurations.
The most important admixture is expected from the $4f^{14} 6s6p ~^3\!P_2$ state because (i) the energy difference, $\Delta E$, between $J=2$ and $^3\!P_2$ is only 3478 cm$^{-1}$ and (ii) even a small admixture of the $^3\!P_2$ state
to $4f^{13} 6s^2 5d ~(J=2)$ leads to an appearance of the permitted $M1$ $6s6p ~^3\!P_2 -\, 6s6p ~^3\!P_1$ transition
and, as a result, can significantly change the $M1$ $(J=2)-\, ^3\!P_1$ transition amplitude.
To reproduce correctly the $(J=2) -\, ^3\!P_2$ mixing, an accurate calculation of $\Delta E$ is required.

We start from a solution of the Dirac-Fock equations and carry out the initial self-consistency procedure for the [$1s^2,..., 4f^{14},6s^2$]
configuration. Then, all electrons were frozen and one electron from the $4f$ shell was moved to the $6p$ shell, so the $6p_{1/2,3/2}$ orbitals were constructed for the $4f^{13} 6s^2 6p$ configuration. The $5d_{3/2,5/2}$ orbitals were constructed for the $4f^{13} 6s^2 5d$ configuration.
The configuration space was formed by allowing single and double excitations for the even-parity states from the configurations $4f^{14} 6s^2$, $4f^{14} 6s5d$, and  $4f^{13} 6s^2 6p$ and for the odd-parity states from the configurations $4f^{14} 6s6p$ and $4f^{13} 6s^2 5d$.
The basis set used in the CI calculations included virtual orbitals up to  $8s$, $8p$, $7d$, $7f$, and $7g$, which we labeled as $8sp7dfg$ CI set. Smaller CI sets were used to check the uncertainty of the calculations.

The CI method  does not reproduce $\Delta E$ with sufficient accuracy leading to an underestimated value for the $M1$ $(J=2)-\, ^3\!P_1$ transition amplitude, and therefore to an overestimated lifetime of the $4f^{13} 6s^2 5d ~(J=2)$ state of about $\sim 22$ min.
To provide a lowest bound on the lifetime, we model correct $\Delta E$ and, therefore, correct mixing,
using a different sensitivity of the $J=2$ and $^3\!P_J$ states to $\alpha$ variation. We changed the fine structure constant by
7\% and repeated the CI calculations as described above. It allowed us to bring these levels together to
$\Delta E \approx 3630\, {\rm cm}^{-1}$, which is very close to the experimental value. This computation was used to establish the lower bound on the lifetime of the $J=2$ level.
The estimates of the $M1$ transitions (in Bohr magnetons $\mu_0$) and $E2$ transitions matrix elements
(in $|e|a_0^2$, where $e$ and $a_0$ are the electron charge and Bohr radius), transition rates A (in s$^{-1}$), and  branching ratios obtained using both calculations are listed in Table~\ref{tab2}.
The experimental energies are used in the calculation of the transition rates.

\begin{table*} [h]
\caption{\label{tab2}
 Magnetic-dipole ($M1$) (in $\mu_0$) and electric-quadrupole ($E2$) (in $|e|a_0^2$) reduced matrix elements (MEs), transition rates A (in s$^{-1}$), and  branching ratios (Br.) for the $4f^{13}6s^25d ~(J=2) -4f^{14}6s6p~^3\!P_J$ transitions.  The lifetime $\tau$ (in s) of the $J=2$ state is given in the last row.
 The results of two computations are presented. The first computation of the matrix elements is carried out using the large-scale \textit{ab initio} CI
calculation of the matrix elements with the $8sp7dfg$  CI set.
The second calculation specifically shifts the splitting of the $J=2$ and $^3\!P_2$
levels to $\Delta E \approx 3630\, {\rm cm}^{-1}$, close to the experimental value to establish the effects of their mixing.
Experimental energies are used in all transition rate calculations. }
\begin{ruledtabular}
\begin{tabular}{lccccccc}
\multicolumn{1}{c}{Transition} &
\multicolumn{1}{c}{} &
\multicolumn{3}{c}{\textit{ab initio} CI calculations} &
\multicolumn{3}{c}{Model calculations}\\
\multicolumn{2}{c}{} &
\multicolumn{1}{c}{ME} &
\multicolumn{1}{c}{$A$}&
\multicolumn{1}{c}{Br.} &
\multicolumn{1}{c}{ME} &
\multicolumn{1}{c}{$A$}&
\multicolumn{1}{c}{Br.} \\
\hline \\[-0.8pc]
$4f^{13}6s^25d ~(J=2) - \,^3\!P_0$ & $E2$  & 0.46  & $ 3.4\times10^{-5}$ & 0.05  & 1.7  & $ 4.6\times10^{-4}$ & 0.03 \\
$4f^{13}6s^25d ~(J=2) - \,^3\!P_1$ & $M1$ & 0.03  & $ 6.6\times10^{-4}$ & 0.88   & 0.15 & $ 1.7\times10^{-2}$ & 0.94 \\
                                   & $E2$  & 0.74  & $ 4.6\times10^{-5}$ & 0.06  & 2.7  & $ 6.2\times10^{-4}$ & 0.03 \\
$4f^{13}6s^25d ~(J=2) - \,^3\!P_2$ & $M1$ & 0.004 & $ 3.0\times10^{-6}$ & 0.00   & 0.02 & $ 4.0\times10^{-6}$ & 0.00 \\
                                   & $E2$  & 0.68  & $ 5.2\times10^{-6}$ & 0.01  & 2.4  & $ 6.6\times10^{-5}$ & 0.00 \\
Lifetime                             &     &       &                    & 1300~s &      &                     & 55~s  \\
\end{tabular}
\end{ruledtabular}
\end{table*}

\end{document}